\documentclass[conference, 9pt]{IEEEtran}\usepackage[compatibility=false]{caption}
\IEEEoverridecommandlockouts
\usepackage{cite}
\usepackage{amsmath,amssymb,amsfonts}
\usepackage{algorithmic}
\usepackage{graphicx}
\usepackage{textcomp}
\usepackage{xcolor}
\usepackage{url}
\usepackage{caption}
\usepackage{subcaption}

\usepackage{stfloats}
\usepackage{siunitx}
\usepackage{makecell}
\usepackage{multirow}
\usepackage{hyperref}
\usepackage{booktabs}
\usepackage[normalem]{ulem}
\usepackage{tabularx}

\usepackage[T1]{fontenc}

\def\BibTeX{{\rm B\kern-.05em{\sc i\kern-.025em b}\kern-.08em
    T\kern-.1667em\lower.7ex\hbox{E}\kern-.125emX}}

\newif\iftodo
\todotrue

\begin{document}

\title{Generative Data Augmentation Challenge:\\ Synthesis of Room Acoustics for Speaker Distance Estimation\thanks{This material is in part based on work supported by the National Science Foundation under grant numbers 2512987 and 1910940.

The authors appreciate Michael Neri at Roma Tre University for the guidance in implementing the baseline distance estimation model, and Anton Ratnarajah for guidance on the GWA dataset.}}
\author{\IEEEauthorblockN{Jackie Lin$^{1,7}$, Georg Götz$^2$, Hermes Sampedro Llopis$^2$, Haukur Hafsteinsson$^2$, Steinar Gu\dh jónsson$^2$, Daniel Gert Nielsen$^2$,\\ Finnur Pind$^2$,  Paris Smaragdis$^1$, Dinesh Manocha$^3$, John Hershey$^4$, Trausti Kristjansson$^{5,6}$, and Minje Kim$^{1,5,8}$}
\IEEEauthorblockA{
$^1$\textit{University of Illinois at Urbana-Champaign}, $^2$\textit{Treble Technologies}, $^3$\textit{University of Maryland},\\$^4$\textit{Google Research}, $^5$\textit{Amazon Lab126}, $^6$\textit{Reykjavik University}\\
$^7$jackiel4@illinois.edu, $^8$minje@illinois.edu}}

\maketitle

\begin{abstract}
This paper describes the synthesis of the room acoustics challenge as a part of the generative data augmentation workshop at ICASSP 2025. The challenge defines a unique generative task that is designed to improve the quantity and diversity of the room impulse responses dataset so that it can be used for spatially sensitive downstream tasks: speaker distance estimation. The challenge identifies the technical difficulty in measuring or simulating many rooms' acoustic characteristics precisely. As a solution, it proposes generative data augmentation as an alternative that can potentially be used to improve various downstream tasks. The challenge website, dataset, and evaluation code are available at \href{https://sites.google.com/view/genda2025}{https://sites.google.com/view/genda2025}.

\end{abstract}

\begin{IEEEkeywords}
Room acoustics, spatial audio, room impulse response, speaker distance estimation 
\end{IEEEkeywords}

\section{Introduction}
In recent years, the room acoustics and spatial audio research field has seen a surge of research motivated by new technology such as augmented and virtual reality (AR/VR), deep learning, and parallel computing. The technology is largely based on precise characterization of the room acoustics by considering various factors, such as the shape of the room, furniture layout, positions and directionality of the sources and receivers, the material of the walls, and more. 

One of the most effective, yet exhaustive characterizations of a room's acoustics is a dense set of room impulse responses (RIR) recorded at fine enough source-receiver positions in the room. Indeed, the explosion of learning-based approaches for old and new room acoustics tasks has increased the demand for such RIR data. However, when the RIR datasets used to train spatial audio systems are not diverse enough, these systems often struggle to generalize to new, unfamiliar rooms. This leads to poor user experiences or inaccurate results. Furthermore, gathering a diverse and large set of RIRs is in fact quite difficult; the process and equipment for measuring high-quality RIRs are costly and technical, and traditional room acoustics simulation methods suffer strong trade-offs between accuracy and computational cost. Over the years, many dataset papers have attempted to fill gaps in the public dataset ecosystem\footnote{An online repository of RIR datasets is maintained at \href{https://github.com/RoyJames/room-impulse-responses}{https://github.com/RoyJames/room-impulse-responses}.}~\cite{jeub2009binaural, eaton2016estimation, traer2016statistics, chen2020soundspaces, koyama2021meshrir, tang2022gwa, gotz2021dataset}, demonstrating the demand for and the challenge of obtaining quality room acoustics data.

\begin{figure}[t]
    \centering
    \includegraphics[width=.86\linewidth]{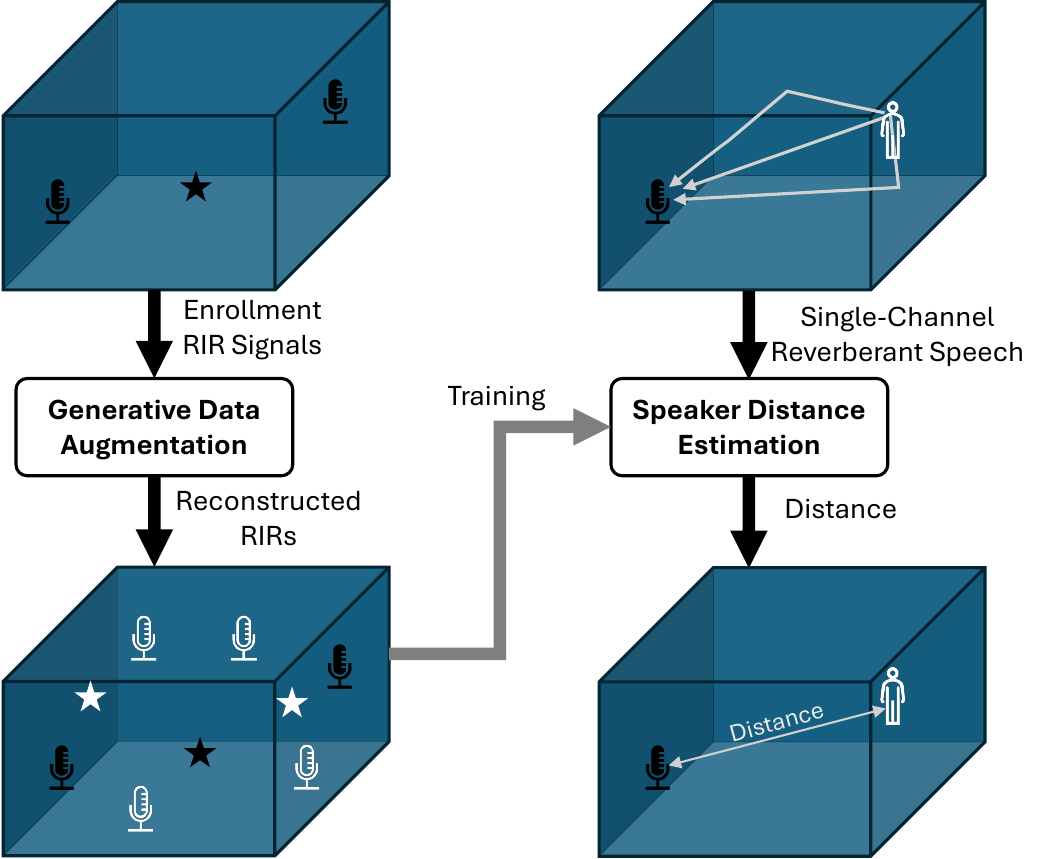}
    \caption{Overview of the generative data augmentation pipeline.}
    \label{fig:overview}
\end{figure}

In this challenge, we recognize the inherent value of generative AI models for data augmentation. As demonstrated in other domain areas such as computer vision~\cite{kirillov2023segment, bao2017cvae-gan, XuY2021cvpr}, speech synthesis~\cite{kuznetsova2023potential, rosenberg2019speech}, and natural language understanding~\cite{li2022data}, a well-designed generative model can help augment existing datasets that are often limited in quantity and diversity. The proposed challenge promotes research in generating RIR datasets. The synthesized RIRs can be directly used for various applications such as room auralization and spatial audio rendering, but this challenge focuses on using them to augment existing RIR datasets to train a better machine learning (ML) model. To this end, we propose to use the augmented dataset to improve the \textit{speaker distance estimation} (SDE) task, where the ML model predicts the physical distance between a speech source and the receiver based only on the single-channel reverberant observation. 

The challenge asks participants to develop a generative model to reconstruct the RIRs of several rooms, given only a limited amount of information about the rooms, such as only a few RIR signals and basic shapes of the rooms. Then, the organizers compare participants’ reconstructed RIRs with the held-out ground-truth RIRs to assess their quality. In addition, participants are asked to train SDE models from a reverberant speech dataset they construct based from the few provided RIR signals. Their distance estimation performance will be evaluated on the hidden test signals, whose actual distances are kept from the participants. The overall pipeline of the generative data augmentation used for SDE is shown in Fig. \ref{fig:overview}.

In this paper, we describe the challenge, datasets, evaluation methods, and baseline models. The enrollment RIRs, 3D models of the rooms, test room-source-receiver locations, reverberant speech test set, evaluation metric functions, and trained baseline SDE model are provided in this link: \href{https://sites.google.com/view/genda2025}{https://sites.google.com/view/genda2025}.

\section{Overview of the Challenge}

The challenge tasks participants to generate a room's acoustics in the form of RIRs using a limited amount of information about the room. The assumption is that the acoustical information of the room is difficult to measure or acquire for a normal user, while knowing the room acoustics is very important for end-user applications, such as AR/VR. Hence, a data augmentation method can be used to flesh out a specific room's acoustic information from an elementary set of available information that a non-technical user can easily provide, such as a few RIRs at a handful of locations. The data augmentation system can figure out other RIRs from various locations in the room to the degree that is useful enough for other downstream tasks. 

The challenge consists of two tasks that evaluate the room acoustics data augmentation on both direct RIR generation and its practical application in a downstream task:


\subsection{Task 1: Evaluating Direct RIR Generation} Participants are asked to generate RIRs at organizer-defined source-receiver locations for each of the rooms to directly evaluate their room acoustics data augmentation system. The organizers will also provide other information for advanced generation methods, such as the 3D model of the room shape, furniture layouts, and Ambisonic RIRs.
The Task 1 submission is generated RIRs at held-out source-receiver locations. A straightforward method to assess the quality of the participating generative system is to directly compare the submission with the corresponding ground truth by using various objective metrics as listed in Sec. \ref{sec:rir_eval}. 

\subsection{Task 2: Evaluating Usefulness in Downstream Task} 
Beyond direct RIR evaluation metrics, the challenge validates generated RIRs through their effectiveness in a downstream task. Ideally, a successful generative AI system could recover RIRs of the room with near-perfect precision. However, by indirectly measuring the quality of the RIRs in the context of the downstream task, it is expected that the generative AI systems can focus more on the properties that are important for the task's specificity. We choose the single-channel Speaker Distance Estimation (SDE) task because it is a challenging problem, relying only on the spectral and temporal characteristics of the observed signal in the single-channel scenario. 

To standardize evaluation and reduce participant workload, the organizers provide a state-of-the-art SDE model as a baseline (detailed in Sec~\ref{sec:sde}). Participants must use this model architecture without modifications, ensuring that comparisons reflect data quality rather than architectural choices.
The task requires participants to:
\begin{enumerate}
    \item Create training data using their data augmentation systems. 
    Participants can generate training data by simulating virtual speakers at various locations within the enrollment rooms. The generated RIRs, with the known source-receiver geometry providing the ground-truth distance labels for training, are to be used to construct the reverberant speech dataset.
    \item Fine-tune the baseline SDE model using only their generated data
\end{enumerate}

While a truly robust SDE model should generalize to any environment, this challenge focuses on improving performance through fine-tuning for specific provided rooms. The key challenge lies in generating RIRs realistic enough for the model to learn accurate room acoustics. The fine-tuned SDE model is evaluated on reverberant speech utterances recorded at undisclosed locations within the challenge dataset. The Task 2 submission is the participants' SDE models' distance estimates, which are compared against the ground truth distances.

The organizers provide a bonus track to highly encourage new model architectures for SDE, which participants can utilize to explore more structural variations.

\begin{table}[t]
\vspace{-6pt}
\centering
\caption{Overview of the rooms in the dataset.}
\label{tab:rooms}
\vspace{2pt}
\resizebox{0.95\columnwidth}{!}{
\begin{tabular}{llcc}
\toprule
\textbf{Room ID} & \textbf{Description} & Measured & Simulated\\
\midrule
Room\_0\hspace{0.5em} & Control - Treble room & \checkmark  & \checkmark\\
\midrule
Room\_1\hspace{1em} & Bathroom 1 && \checkmark\\
Room\_2\hspace{1em} & Bathroom 2 && \checkmark\\
Room\_3\hspace{1em} & Bedroom 1 && \checkmark\\
Room\_4\hspace{1em} & Bedroom 2 && \checkmark\\
Room\_5\hspace{1em} & Living room with hallway 1 && \checkmark\\
Room\_6\hspace{1em} & Living room with hallway 2 && \checkmark\\
Room\_7\hspace{1em} & Living room 1 && \checkmark\\
Room\_8\hspace{1em} & Living room 2 && \checkmark\\
Room\_9\hspace{1em} & Meeting room 1 && \checkmark\\
Room\_10\hspace{1em} & Meeting room 2 && \checkmark\\
Room\_11-20\hspace{1em} & GWA rooms && \checkmark\\
\bottomrule
\end{tabular}
}
\end{table}

\begin{figure}
     \centering
     \begin{subfigure}[b]{0.45\columnwidth}
         \centering
         \includegraphics[width=.88\linewidth]{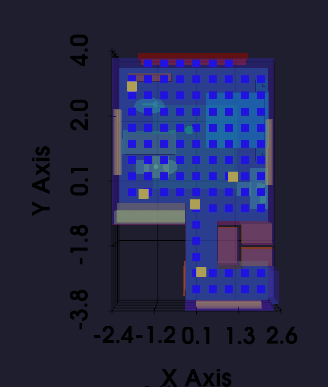}
    \caption{The 3D model of a simulated room. }
         \label{fig:simulatedroom}
     \end{subfigure}
     \begin{subfigure}[b]{0.535\columnwidth}
         \centering
         \includegraphics[width=.9\linewidth]{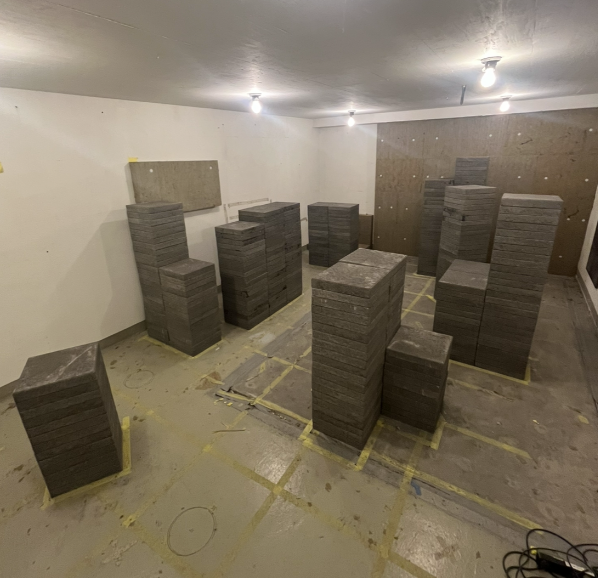}
    \caption{Photograph of measured room `Room\_0' at Treble headquarters.}
         \label{fig:measuredroom}
     \end{subfigure}
     \caption{Images of the Treble simulated rooms.} 
\end{figure}

\section{Datasets}

\subsection{Overview of the Dataset}

The challenge presupposes that wave-based and hybrid wave-geometric room acoustics simulations provide accurate enough room acoustics. Thus, instead of collecting data from real-world rooms, we make our RIR dataset from ten simulated rooms from Treble Technology's software and ten rooms from the GWA dataset~\cite{tang2022gwa}. The twenty rooms have varying shapes, furniture layouts, and wall absorption characteristics. 

Treble Technologies' room simulation software accurately models room acoustics, as evidenced by multiple prior validation studies~\cite{TrebleTechnologies}. We use the Treble SDK, a Python interface to the wave-based solver, thus enabling the efficient setup and simulation of large RIR datasets. The wave-based solver uses the Discontinuous Galerkin~(DG) method~\cite{pind2020time, melander2024massively}. Although the Treble software also allows for a hybrid combination with a Geometrical Acoustics solver, the rooms were simulated purely with a wave-based solver to ensure the best accuracy. The largest simulated frequency was \SI{7}{\kilo\Hz}. 

Ten more rooms from the GWA dataset, which are simulated using a hybrid wave FDTD and ray-tracing method, are included in the challenge dataset. This increases the diversity of RIRs available to participants.

\noindent\textbf{Simulated RIRs}: The Treble simulated portion of the RIR dataset consists of 3085 simulated monaural and 8th order Ambisonics spatial RIRs at \SI{32}{\kilo\Hz} sampling rate with labeled source-receiver positions. This simulated dataset consists of ten rooms that contain furniture and have assigned surface materials. A description of the rooms is provided in Table~\ref{tab:rooms}, and an example room is shown in Fig.~\ref{fig:simulatedroom}. For each room, RIRs are simulated for five source locations and many microphone positions in a grid with \SI{0.5}{\meter} spacing at $\{0.5, 1.0, 1.5\}$m elevation. Any mics that collide with room or furniture geometry are omitted. The rooms dimensions range from \SI{1.75}{\meter} to \SI{6.3}{\meter} and the number of microphones per room ranges from 14 to 137. 

The GWA portion of the RIR dataset consists of 1026 simulated monaural RIRs at \SI{48}{\kilo\Hz} sampling rate with labeled source-receiver positions. Like the Treble rooms, the GWA rooms also contain rich room and furniture layouts. In all, the dataset has a total of 4111 RIRs.

\noindent\textbf{Control room}: 
The organizers also provide a control set of real-world room recordings and their counterpart simulations (`Room\_0' in Table~\ref{tab:rooms} \& Fig.~\ref{fig:measuredroom}). The measured room, a real physical room in Treble's headquarters, contains a complicated furniture layout and wall absorption panels. Two loudspeaker positions and ten mic positions are recorded, yielding 20 RIRs with labeled source-receiver positions. Meanwhile, the simulation replicates the real room, `Room\_0', using the same Treble simulation software, while additionally simulating a \SI{0.5}{\meter} grid of RIRs. This paired data is provided for participants to calibrate their models, which could be biased toward the different datasets they were trained on. 

\subsection{Enrollment Dataset} \label{sec:scenario}

The organizers withhold a majority of the dataset from participants, as this challenge asks participants to generate RIRs from limited room information. We are interested in replicating two real-world user scenarios where only sparse room measurements are available, from which participants of this challenge must generate unseen source-receiver position RIRs. The two scenarios help us choose a subset of our simulated dataset to provide as enrollment data and help us choose the test data to effectively evaluate the participating generative RIR systems (ref. Table~\ref{tab:task}). The scenarios, enrollment data, and requested distance estimations of test speech are summarized in Table~\ref{tab:task}.

\noindent\textbf{Scenario 1 -- Center-to-corner augmentation}: Ten rooms are chosen to represent the case when the enrollment signals are of sources recorded by one or more mics near the center location of the room. For each room in Scenario 1, five or more enrollment RIRs are given. 

\noindent\textbf{Scenario 2 -- Corner-to-corner augmentation}: The other ten rooms are for the case when the enrollment RIRs are recorded from source-receiver positions that are close to the corners of the room. The rest of the process is the same, but for this case, the RIR generation is challenged in a different way as it is exposed only to partial information from a handful of corner positions. 

The enrollment data, representing a realistic sparse subset of the overall dataset, are provided. The monoaural RIRs, 8th order Ambisonics RIRs, and source and receiver positions of the Treble rooms are given. For the GWA rooms, only the monaural RIRs and source-receiver positions are given. The 3D models of the Treble rooms are also provided and can be used in place of or in addition to the enrollment RIRs to encourage participation from participants who work on applications where the 3D room geometry is obtainable. Although the simulations incorporated material properties, this information is not disclosed to the participants. Instead, the 3D geometry is organized into layers with semantic labels (e.g., ``window", ``interior wall",  ``chair," etc.).

\begin{table}[t]
\footnotesize
\resizebox{\columnwidth}{!}{
\begin{tabularx}{\columnwidth}{@{}
    >{\raggedright\arraybackslash}X@{\hspace{-10pt}}
    >{\raggedright\arraybackslash}X@{\hspace{-10pt}}
    >{\centering\arraybackslash}X@{\hspace{-5pt}}
    >{\centering\arraybackslash}X@{\hspace{-5pt}}
    >{\centering\arraybackslash}X@{}}
\toprule
\textbf{Scenario} & \multicolumn{2}{c}{\textbf{Enrollment Data}} & \multicolumn{2}{c}{\textbf{Evaluation}}\\
 & Rooms & \# Enrollment & Task 1 & Task 2\\
& & RIRs & \# Generated RIRs & \# Distance Predictions\\
\midrule
Scenario 1 & 1, 3, 5, 7, 9, 11, 13, 15, 17, 19 & 59 & 100 & 240\\
Scenario 2 & 2, 4, 6, 8, 10, 12, 14, 16, 18, 20 & 57 & 100 & 240\\
\midrule
\textbf{Total} & 20 & 116 & \textbf{200} & \textbf{480}\\
\bottomrule
\end{tabularx}
}
\caption{\textbf{Overview of Scenarios, Enrollment Data, Evaluation RIRs, and Test Speech.} For each scenario (see Sec.~\ref{sec:scenario} for descriptions), participants are given enrollment RIRs. In Task 1 they must generate 10 RIRs per room at specified unseen source-receiver positions, and in Task 2 predict 480 speaker distances.}
\label{tab:task}
\end{table}

\begin{table*}[t]
    \centering
    \resizebox{.9\textwidth}{!}{  \begin{tabular}{l|c|cccccc|c|cccccc|c}
        \toprule
         & \multicolumn{7}{c|}{$\bar\tau_{20}$ \textbf{T20 MAPE} [\%]} & \multicolumn{7}{c|}{$\bar\omega$ \textbf{EDF MSE} [dB]} & \multirow{2}{*}{ $\bar\rho$ {\textbf{DRR MSE} [dB]}} \\ 
        & {Full} & {125} & {250} & {500} & {1k} & {2k} & {4k} & {Full} & {125} & {250} & {500} & {1k} & {2k} & {4k} & \\ \cline{1-1}
        \midrule
        \makecell{\textbf{Treble Simulation}} & 13.5 & 10.0 & 14.9 & 10.9 & 12.1 & 6.9 & 17.8 & 22 & 7.0 & 7.6 & 5.0 & 4.4 & 7.7 & 41.1 & 14.5 \\
        \bottomrule
    \end{tabular}}
    \caption{\textbf{Quantitative Error of the Treble Simulated vs. Measured Room}. Broadband and octave-band T20 MAPE and EDF MSE, and DRR MSE of the simulated RIRs compared to the measured RIRs in Treble's room. The low errors indicate thatthe rooms simulated with Treble software in our challenge dataset matches real-world data.}
    \label{tab:measuredvsim}
\end{table*} 


\section{Task 1: RIR Generation}\label{sec:rir_eval}
\subsection{Generated RIR Evaluation}

Participants are asked to submit ten generated RIRs from each of the twenty rooms for a total of 200 RIR signals from the designated source-receiver locations. Participants' generated RIRs will be evaluated against the hidden subset of challenge RIRs on the following commonly used metrics:

\noindent\textbf{T20 Mean Absolute Percentage Error}: Reverberation time (T20) is the time it takes for energy in a room to decay $\SI{60}{\decibel}$, obtained from the 20dB evaluation range, as calculated in~\cite{ISO3382-2_2008}. Reverberation time is a key RIR metric and characterizes the perceived reverberance of a space. The T20 mean absolute percentage error (MAPE) $\bar{\tau}_{20}$ is defined as follows:
\begin{equation}
    \bar{\tau}_{20} = \frac{1}{N} \sum_{i=1}^{N} \frac{ |\hat{\tau}_{20}^{(i)} - {\tau}_{20}^{(i)} | } {{\tau}_{20}^{(i)}}
\end{equation}
    where ${\tau}_{20}^{(i)}$ and $\hat{\tau}_{20}^{(i)}$ are the $i$-th ground-truth T20 value and that of the predicted RIR's, respectively. Participants will be evaluated on the broadband T20 MAPE and octave-band T20 MAPEs with center frequencies $f_c = \{125, 250, 500, 1000, 2000, 4000\}$.

\noindent\textbf{EDF Mean Squared Error}: The energy decay relief (EDF) is the set of spectral decay signatures calculated by Schroeder's backwards integration method~\cite{schroeder1965new}. Participants will be evaluated on the mean squared error (MSE) of the broadband and octave-band EDFs. The MSE is calculated on the EDF with the last 5\% of samples discarded, as in~\cite{gotz2021dataset}. This spectral decay signature offers more fine-grained information than the T20. The broadband EDF MSE $\bar\omega$ is
\begin{equation}
    \bar\omega = \frac{1}{N} \sum_{i=1}^{N}{| \hat\omega_{i} - \omega_{i}|^2} \quad 
\end{equation}
and the octave-band EDF MSE is
\begin{equation}
    \bar\omega_{j} = \frac{1}{N} \sum_{i=1}^{N} {| \hat\omega_{ji} - \omega_{ji}|^2} \quad 
\end{equation}
where $\omega_{ji}$ is the energy decay function of octave band $j$ of RIR $i$.

\noindent\textbf{DRR Mean Squared Error}. Direct to reverberant ratio (DRR) is the ratio in $\SI{}{\decibel}$ between the direct sound energy and the rest of the RIR, as calculated in~\cite{ISO3382-2_2008}. The DRR MSE $\bar\rho$ is
\begin{equation}
    \bar\rho = \frac{1}{N} \sum_{i=1}^{N}{(\hat\rho_{i} - \rho_{i})^2} \quad .
\end{equation}

\renewcommand{\arraystretch}{1}
\begin{table*}[t]
    \centering
    \resizebox{\textwidth}{!}{  
    \begin{tabular}{l|l|cc|cc|cc|cc|cc}
    \toprule
    & & \multicolumn{2}{c|}{\textbf{Overall}} & \multicolumn{2}{c|}{\textbf{0--2m}} & \multicolumn{2}{c|}{\textbf{2--4m}}  & \multicolumn{2}{c|}{\textbf{4--6m}} & \multicolumn{2}{c}{\textbf{6+m}}\\ 
    Scenario & Model & MAE [m] & MAPE [\%]&  MAE [m] & MAPE [\%]& MAE [m] & MAPE [\%]& MAE [m] & MAPE [\%] & MAE [m] & MAPE [\%]\\ 
     \midrule
    Scenario 1 & \textbf{SDE\_scenario\_1\_Oracle} & 0.208 & 6.3\% & 0.089 & 7.1\% & 0.207 & 6.9\% & 0.157 & 3.4\% & 0.867 & 10.7\%\\ 
     & \textbf{SDE\_C4DM (Baseline)} & 1.65 & 106\% & 3.06 & 290\% & 0.973 & 35.9\% & 0.804 & 16.7\% & 2.22 & 29.2\% \\ 
    \midrule
    Scenario\ 2 & \textbf{SDE\_scenario\_2\_Oracle} & 0.209 & 7.5\% & 0.123 & 9.2\% & 0.200 & 6.7\% & 0.269 & 5.5\% & 0.524 & 8.2\% \\ 
     & \textbf{SDE\_C4DM (Baseline)} & 1.69 & 101\% & 2.91 & 238\% & 0.977 & 36.1\% & 1.15 & 23.6\% & 1.42 & 22.4\% \\ 
    \bottomrule
    \end{tabular}}
    \caption{Speaker Distance Estimation Error - SDE models trained on the full challenge dataset versus the C4DM dataset show that the oracle systems perform much better on the challenge dataset set than the baseline system which only saw C4DM RIRs.}
    \label{tab:sde1}
\end{table*}


We demonstrate that the Treble simulator generates accurate RIRs and that the simulated dataset is legitimate. To establish a performance upper bound, we evaluated the simulated RIRs in the control room 'Room\_0' against the corresponding measured RIRs provided by Treble Technologies. The results are shown in Table~\ref{tab:measuredvsim}. We set Treble's wave-based simulations as the upper bound of the performance of generative RIR systems. Furthermore, since the simulated dataset was generated using ground-truth material properties of the real room, it should achieve higher accuracy than participants who are not provided with the material data. 

\section{Task 2: Speaker Distance Estimation} \label{sec:sde}

\subsection{Overview of the SDE Task}

Beyond the typical RIR evaluation metrics listed in Sec. \ref{sec:rir_eval}, which are primarily related to the perceptual qualities of RIRs, depending on the application it may be more important that a generative AI system generates good-quality data in the context of a downstream task. To that end, we implement a learning-based speaker distance estimation (SDE) model based on a state-of-the-art open-source model~\cite{neri2024speaker} to test the quality of RIRs generated directly for the SDE task. 

Given the enrollment data, participants are asked to generate as many RIRs as they need to build an RIR dataset to train an SDE model. They are required to fine-tune the baseline model the organizers provide. Note that we fix the model architecture for all participants to restrict the comparison to the quality of the data rather than the model architecture. Also, participants are not allowed to use any RIR data that is not generated from their generative system. Again, this challenge evaluates the quality of the RIR generation system based on the generated dataset's usefulness in improving SDE, so it would be uninformative if the SDE model were improved with other RIR data. 

\noindent\textbf{Bonus Track -- Unconstrained SDE Model}: Lastly, participants are allowed to develop a new SDE model architecture different from the provided baseline and can use any RIR data for training. The bonus models are evaluated on the same reverberant speech test set.

\subsection{The Baseline SDE Models and Their Performance} \label{sec:oracle}

The challenge organizers develop a baseline model and two oracle models for participants to gauge the lower and upper bounds respectively, of SDE model they develop. To this end, as a lower bound baseline, the challenge organizers first adopt the state-of-the-art SDE model provided by the authors of~\cite{neri2024speaker}. This SDE model is a convolutional recurrent neural network with an attention module. It consists of attention masks that prioritize sections of the input recording, convolutional layers, followed by gated recurrent units (GRUs). The model's input is a spectrogram $\{\text{fft\_len}=1024, \text{hop\_size}=512\}$ of 10 seconds, \SI{32}{\kilo\Hz} reverberant speech and the output is a distance in meters. The baseline model is trained on reverberant speech constructed from measured RIRs from the C4DM~\cite{stewart2010database} dataset and the VCTK-Corpus speech dataset~\cite{Yamagishi_Veaux_MacDonald_2019}. It has not seen the particular rooms the challenge introduces, and the checkpoint of this baseline model is provided to participants as a starting point.

As for the oracle case, two SDE models are trained from scratch using the challenge dataset. In doing so, the combined training and validation set of the oracle models consists of all the RIRs excluding the ones used for the test sets, thus ensuring the training process utilizes most of the the dataset. Since this amount of RIR data is hard to acquire in practice, we can consider it as oracle performance. The final train-valid-test split of the challenge dataset is \{3829, 202, 80\} for both scenarios 1 and 2. During each training epoch, each RIR is convolved with a randomly sampled speech recording from a training set of VCTK speakers. Following the hyperparameters provided in~\cite{neri2024speaker}, the model is trained for 50 epochs with a batch size of 16 and an initial learning rate of 0.001. The training took approximately fifteen minutes on two NVIDIA A1000 GPUs. 

The baseline and oracle models are evaluated on the test set which consists of 8 RIRs per room convolved with three randomly sampled speech recordings from a test set of VCTK speakers, totaling $[20 \text{ rooms}]\times[8 \text{ positions}]\times[3 \text{ speakers}]=480$ reverberant speech utterances. Participants are also asked to submit distance predictions on this test set, and lower distance estimation error indicates better performance of the generative RIR system.

The evaluation metric of the SDE model is the distance MAE and the MAPE, which is the absolute distance error as a percentage of the ground truth distance. The results are shown in Tab.~\ref{tab:sde1}. The oracle SDE models trained on scenario 1 and scenario 2 perform well, achieving on average only \SI{20.8}{\centi\meter} and \SI{20.9}{\centi\meter} MAE and 6.3\% and 7.5\% MAPE respectively. When evaluating the test set on the baseline model trained only on the C4DM data, we can see the error increase to \SI{1.65}{\meter} MAE and 106\% MAPE in scenario 1 and \SI{1.69}{\meter} MAE and 101\% MAPE. This is likely due to the test set skewing towards short distances, which are out of the C4DM distance distribution, which consists of measurements from three large spaces.

\section{Conclusion}
In this paper, we propose a room acoustics generative AI challenge that rethinks the potential of generative data augmentation in the domain of spatial audio. First, we recognize the long-standing challenges and work in collecting high-quality and diverse RIRs. Furthermore, we propose the task of augmenting, thereby multiplying, sparse existing acoustic data for fine-tuning speaker distance estimation models. The challenge demonstrates how generative models can enhance the generalization capabilities of SDE models, and proposes one conception of a generative AI framework for developing more robust and better performing spatial audio technologies.

\bibliographystyle{ieeetr}
\bibliography{references}

\begin{thebibliography}{10}

\bibitem{jeub2009binaural}
M.~Jeub, M.~Schafer, and P.~Vary, ``A binaural room impulse response database for the evaluation of dereverberation algorithms,'' in {\em 2009 16th International Conference on Digital Signal Processing}, pp.~1--5, IEEE, 2009.

\bibitem{eaton2016estimation}
J.~Eaton, N.~D. Gaubitch, A.~H. Moore, and P.~A. Naylor, ``Estimation of room acoustic parameters: The {ACE} challenge,'' {\em IEEE/ACM Transactions on Audio, Speech, and Language Processing}, vol.~24, no.~10, pp.~1681--1693, 2016.

\bibitem{traer2016statistics}
J.~Traer and J.~H. McDermott, ``Statistics of natural reverberation enable perceptual separation of sound and space,'' {\em Proceedings of the National Academy of Sciences}, vol.~113, no.~48, pp.~E7856--E7865, 2016.

\bibitem{chen2020soundspaces}
C.~Chen, U.~Jain, C.~Schissler, S.~V.~A. Gari, Z.~Al-Halah, V.~K. Ithapu, P.~Robinson, and K.~Grauman, ``Soundspaces: Audio-visual navigation in 3d environments,'' in {\em Computer Vision--ECCV 2020: 16th European Conference, Glasgow, UK, August 23--28, 2020, Proceedings, Part VI 16}, pp.~17--36, Springer, 2020.

\bibitem{koyama2021meshrir}
S.~Koyama, T.~Nishida, K.~Kimura, T.~Abe, N.~Ueno, and J.~Brunnstr{\"o}m, ``{MeshRIR}: A dataset of room impulse responses on meshed grid points for evaluating sound field analysis and synthesis methods,'' in {\em 2021 IEEE workshop on applications of signal processing to audio and acoustics (WASPAA)}, pp.~1--5, IEEE, 2021.

\bibitem{tang2022gwa}
Z.~Tang, R.~Aralikatti, A.~J. Ratnarajah, and D.~Manocha, ``{GWA}: A large high-quality acoustic dataset for audio processing,'' in {\em ACM SIGGRAPH 2022 Conference Proceedings}, pp.~1--9, 2022.

\bibitem{gotz2021dataset}
G.~G{\"o}tz, S.~J. Schlecht, and V.~Pulkki, ``A dataset of higher-order ambisonic room impulse responses and 3d models measured in a room with varying furniture,'' in {\em 2021 Immersive and 3D Audio: from Architecture to Automotive (I3DA)}, pp.~1--8, IEEE, 2021.

\bibitem{kirillov2023segment}
A.~Kirillov, E.~Mintun, N.~Ravi, H.~Mao, C.~Rolland, L.~Gustafson, T.~Xiao, S.~Whitehead, A.~C. Berg, W.-Y. Lo, {\em et~al.}, ``Segment anything,'' in {\em Proceedings of the IEEE/CVF International Conference on Computer Vision}, pp.~4015--4026, 2023.

\bibitem{bao2017cvae-gan}
J.~Bao, D.~Chen, F.~Wen, H.~Li, and G.~Hua, ``{{CVAE-GAN}: Fine-Grained Image Generation through Asymmetric Training},'' in {\em IEEE International Conference on Computer Vision (ICCV)}, pp.~2764--2773, 2017.

\bibitem{XuY2021cvpr}
Y.~Xu, Y.~Shen, J.~Zhu, C.~Yang, and B.~Zhou, ``{Generative Hierarchical Features from Synthesizing Images},'' in {\em IEEE/CVF Conference on Computer Vision and Pattern Recognition (CVPR)}, pp.~4430--4430, 2021.

\bibitem{kuznetsova2023potential}
A.~Kuznetsova, A.~Sivaraman, and M.~Kim, ``The potential of neural speech synthesis-based data augmentation for personalized speech enhancement,'' in {\em ICASSP 2023-2023 IEEE International Conference on Acoustics, Speech and Signal Processing (ICASSP)}, pp.~1--5, IEEE, 2023.

\bibitem{rosenberg2019speech}
A.~Rosenberg, Y.~Zhang, B.~Ramabhadran, Y.~Jia, P.~Moreno, Y.~Wu, and Z.~Wu, ``Speech recognition with augmented synthesized speech,'' in {\em 2019 IEEE automatic speech recognition and understanding workshop (ASRU)}, pp.~996--1002, IEEE, 2019.

\bibitem{li2022data}
B.~Li, Y.~Hou, and W.~Che, ``Data augmentation approaches in natural language processing: A survey,'' {\em Ai Open}, vol.~3, pp.~71--90, 2022.

\bibitem{TrebleTechnologies}
``Treble technologies.'' \url{https://docs.treble.tech/validation}.

\bibitem{pind2020time}
F.~Pind, C.-H. Jeong, A.~P. Engsig-Karup, J.~S. Hesthaven, and J.~Str{\o}mann-Andersen, ``Time-domain room acoustic simulations with extended-reacting porous absorbers using the discontinuous {Galerkin} method,'' {\em The Journal of the Acoustical Society of America}, vol.~148, no.~5, pp.~2851--2863, 2020.

\bibitem{melander2024massively}
A.~Melander, E.~Str{\o}m, F.~Pind, A.~P. Engsig-Karup, C.-H. Jeong, T.~Warburton, N.~Chalmers, and J.~S. Hesthaven, ``Massively parallel nodal discontinous {Galerkin} finite element method simulator for room acoustics,'' {\em The International Journal of High Performance Computing Applications}, vol.~38, no.~3, pp.~154--174, 2024.

\bibitem{ISO3382-2_2008}
I.~3382-2, {\em Acoustics—Measurement of room acoustic parameters— Part 2: Reverberation time in ordinary rooms}.
\newblock International Organization for Standardization, 2008.

\bibitem{schroeder1965new}
M.~R. Schroeder, ``New method of measuring reverberation time,'' {\em The Journal of the Acoustical Society of America}, vol.~37, no.~6\_Supplement, pp.~1187--1188, 1965.

\bibitem{neri2024speaker}
M.~Neri, A.~Politis, D.~Krause, M.~Carli, and T.~Virtanen, ``Speaker distance estimation in enclosures from single-channel audio,'' {\em IEEE/ACM Transactions on Audio, Speech, and Language Processing}, 2024.

\bibitem{stewart2010database}
R.~Stewart and M.~Sandler, ``Database of omnidirectional and {B-}format room impulse responses,'' in {\em 2010 IEEE International Conference on Acoustics, Speech and Signal Processing}, pp.~165--168, IEEE, 2010.

\bibitem{Yamagishi_Veaux_MacDonald_2019}
J.~Yamagishi, C.~Veaux, and K.~MacDonald, ``{CSTR VCTK} {Corpus}: {English} multi-speaker corpus for {CSTR} {Voice} {Cloning} {Toolkit} (version 0.92),'' Nov 2019.

\end{thebibliography}

\end{document}